**Above-room-temperature ferromagnetism in ultrathin van der Waals magnet**

Hang Chen[1], Shahidul Asif [2], Kapildeb Dolui[3], Yang Wang[1], Jeyson Tamara Isaza[1,4], V. M. L. Durga Prasad Goli[1], Matthew Whalen[1], Xinhao Wang[1], Zhijie Chen[5], Huiqin Zhang[6], Kai Liu[5], Deep Jariwala[6], M. Benjamin Jungfleisch[1], Chitraleema Chakraborty[1,2], Andrew F. May[7], Michael A. McGuire[7], Branislav K. Nikolic[1], John Q. Xiao[1], Mark J.H. Ku[1,2,†]

[1]*Department of Physics and Astronomy, University of Delaware, Newark, Delaware 19716, United States*
[2]*Department of Materials Science and Engineering, University of Delaware, Newark, Delaware 19716, United States*
[3]*Lomare Technologies Ltd., 6 London Street, London EC3R 7LP, United Kingdom*
[4]*Departamento de Física, Universidad Nacional de Colombia, Bogotá D.C., Colombia*
[5]*Department of Physics, Georgetown University, Washington, DC 20057, United States*
[6]*Department of Electrical and Systems Engineering, University of Pennsylvania, Philadelphia, Pennsylvania 19104, United States*
[7]*Materials Science and Technology Division, Oak Ridge National Laboratory, Oak Ridge, Tennessee 37831, United States*

† Email: mku@udel.edu

Two-dimensional (2D) magnetic van der Waals materials provide a powerful platform for studying fundamental physics of low-dimensional magnetism, engineering novel magnetic phases, and enabling ultrathin and highly tunable spintronic devices. To realize high quality and practical devices for such applications, there is a critical need for robust 2D magnets with ordering temperatures above room temperature that can be created via exfoliation. Here the study of exfoliated flakes of cobalt substituted $Fe_5GeTe_2$ (CFGT) exhibiting magnetism above room temperature is reported. Via quantum magnetic imaging with nitrogen-vacancy centers in diamond, ferromagnetism at room temperature was observed in CFGT flakes as thin as 16 nm. This corresponds to one of the thinnest room-temperature 2D magnet flakes exfoliated from robust single crystals, reaching a thickness relevant to practical spintronic applications. The Curie temperature $T_c$ of CFGT ranges from 310 K in the thinnest flake studied to 328 K in the bulk. To investigate the prospect of high-temperature monolayer ferromagnetism, Monte Carlo calculations were performed which predicted a high value of $T_c$ ~270 K in CFGT monolayers. Pathways towards further enhancing monolayer $T_c$ are discussed. These results support CFGT as a promising platform to realize high-quality room-temperature 2D magnet devices.

**1. Introduction**

The discovery of ultrathin van der Waals (vdW) materials with magnetic ordering [1-6] has opened up enormous opportunities for studying novel condensed matter phenomena and for meeting the demands of new information processing capabilities. Also known as two-dimensional (2D) magnets, they embody a high degree of tunability (e.g., via gating or strain), the ability to form custom-designed, multifunctional heterostructures, and the availability of a rich array of interfacial phenomena. These unique properties position 2D magnets as excellent platforms for fundamental study of low-dimensional spin physics[7], engineering novel quantum phase via proximity effects[6] or Moire structures[8-10], and studying complex magnon behaviors[11,12]. The potential to realize these novel effects, together with the ability for highly efficient interfacial control[13], present 2D magnets as promising platforms to implement a wide range of important applications from spintronics to magnon-based hybrid quantum systems[14,15].

To realize high quality and practical devices for aforementioned applications, achieving a Curie temperature $T_c$ above room-temperature presents a critical goal[4]. $T_c$ significantly below room-temperature is common among 2D magnets; for example, $T_c$ ~60 K in monolayer $CrI_3$[1] and $CrGeTe_3$[2], and ~130 K in monolayer $Fe_3GeTe_2$[16]. Recently, several works have reported magnetism at room-temperature or above in 2D magnetic flakes grown epitaxially, via chemical vapor deposition, or via chemical exfoliation[17-20]. However, materials prepared in such a way are often prone to disorder and are generally unsuitable for creating high quality devices. To date, it is recognized that mechanical exfoliation from bulk material remains the technique of choice to achieve the highest quality in 2D devices. This approach also enables newly discovered bulk materials to be screened rapidly. Among exfoliated flakes, ionically intercalated few-layer $Fe_3GeTe_2$ can achieve $T_c$ above room temperature[21], but likewise does not present a viable path for high quality devices. The possibility of intrinsic room-temperature ferromagnetism in exfoliated ultrathin flakes is recently demonstrated in the 1T phase of $CrTe_2$[22,23] and 2H phase of $VSe_2$[24]. Following these works, there is an urgent need to expand the portfolio of such materials, which is further highlighted by potential robustness issues in the two aforementioned existing vdW systems. In the case of $VSe_2$, its bulk crystal is not magnetic, but a structural phase transition occurs upon thinning and this leads to magnetic thin flakes[24]. However, it is unknown whether this structural transition is sensitive to the substrate or to strain, which may present problems for heterostructure assembly and for creating devices. The magnetic 1T-phase of $CrTe_2$ is meta-stable and decomposes above 330 K[22], which also presents an obstacle for heterostructure/device fabrication which generally requires heating above this temperature. Therefore, ultrathin 2D magnets with $T_c$ above room temperature created via exfoliation of robust crystals present a highly sought-after goal.

In this work, we report above-room-temperature ferromagnetism in exfoliated flakes of Co-substituted $Fe_5GeTe_2$ (CFGT). The family of iron-based vdW metals $Fe_nGeTe_2$ provides a promising platform for engineering high $T_c$ 2D magnet, where $T_c$ can be raised by increasing Fe content. For ultrathin flakes (≲20 nm), $T_c$ is enhanced from ~ 130 K in $Fe_3GeTe_2$[16], to 270 K in $Fe_4GeTe_2$[25] and 270-310 K in $Fe_5GeTe_2$[26,27]. While $Fe_5GeTe_2$ system has a stacking related transition near 570K, the material appears to be thermally and structurally stable[28]. Cobalt substitution is found to further increase $T_c$ in bulk crystal of $Fe_5GeTe_2$[29,30], indicating the potential for above-room-temperature magnetism in ultrathin CFGT flakes. However, thin flakes generally have lower $T_c$ compared to bulk crystals, hence the question of whether ultrathin CFGT flakes could exhibit room-temperature magnetism remains to be explored. We address this question by detecting the minute magnetic stray field signal of CFGT flakes in air at room-temperature and above with a novel quantum magnetic imaging (QMI) technique based on nitrogen-vacancy (NV) centers in diamond. We have observed ferromagnetism at room-temperature in flakes as thin as 16 nm (16 layers). This corresponds to one of the thinnest room-temperature 2D magnet flakes exfoliated from robust single crystals. The measured Curie temperature of CFGT ranges from 310 K in the thinnest flake studied to 328 K in the bulk. To investigate the prospect of high-temperature monolayer ferromagnetism, Monte Carlo simulations using *ab initio* effective spin Hamiltonian have been employed to determine a nearly-room-temperature $T_c$ ~270 K in CFGT monolayer. We discuss the prospect for further enhancing $T_c$ in monolayer CFGT. Our results point to CFGT as a promising platform to realize high-quality devices operational at room-temperature for applications and fundamental studies.

## 2. Results and discussion

### 2.1. Characterization of bulk CFGT crystal

CFGT crystal growth and structural information (Inset in **Figure 1**a) are reported previously[29]. Crystals described in Ref. [29] remain ferromagnetic and have an enhanced $T_c$ ≈330 K up to ≈30% Co-substitution; above this level of substitution, e.g. at ≈50%, the material becomes antiferromagnetic. We note that another work reports the realization of a polar ferromagnetic metal with 50% Co-substituted $Fe_5GeTe_2$[31]. Difference between these two works may be due to minor differences in composition and the resulting structure. Material reported in Ref. [31] also exhibits a high $T_c \approx 350$ K, but magnetic properties were only examined in flakes with thickness >100 nm. In the present work, all CFGT corresponds to the 28% Co-substituted crystals as reported in Ref. [29]. Temperature-dependent magnetization measurements (**Figure 1**a) were performed with a vibrating sample magnetometer and reveals $T_c = 328$ K (for details on the extraction, refer to **S4** in **Supporting Information**) for the bulk crystal, consistent with previous results[29]. Isothermal magnetization data with demagnetization effect accounted for (see **Section S5** in **Supporting Information** for details) is shown in **Figure 1**b which reveals easy-plane anisotropy, likewise consistent with the previous result[29]. Furthermore, we extracted the effective anisotropic field $\mu_0 H_a = -0.51$ T (for details on the extraction, refer to **Section S5** in **Supporting Information**), which has not been reported previously. A zoom-in view of the isothermal magnetization curve (Inset in **Figure 1**b) reveals zero remanence, indicating the presence of magnetic domains in the material at low field that averages to zero magnetization.

## 2.2. Revealing room-temperature magnetism of CFGT flakes with QMI

Given the small dimension and low mass of exfoliated flakes, a sensitive probe of magnetism capable of room temperature operation is required to investigate ultrathin CFGT flakes. To meet this need, we employed wide-field QMI with NV centers in diamond. Given its ability to detect magnetic field with high-sensitivity and high spatial resolution, the NV center in diamond realizes a powerful quantum sensor of local magnetic field and hence provides an enabling tool for advancing condensed matter physics and materials science[32]. In particular, the implementation of NV sensing in the form of wide-field magnetic imaging has been applied to study novel electrical flow in graphene[33], vdW superconductors[34], as well as 2D magnets both at low-temperatures[35] as well as at room-temperature[27]. Hence, QMI provides a highly suitable tool for our investigation.

The setup of QMI is described previously[27] and summarized in **Figure 1**c. An antenna for delivering microwave (MW) necessary for NV measurement is fabricated on the diamond. We exfoliated CFGT flakes on diamond containing a near-surface NV ensemble. We evaporated a thin layer of Al (~5 nm) immediately after exfoliation to protect the flakes from oxidation. As in Ref. [27], optically detected magnetic resonance (ODMR) was performed with a camera to image the stray magnetic field generated by CFGT flakes. We adopted the configuration where an out-of-plane (*z*-direction) bias field $B_0$ is applied, which enables us to image the distribution of the out-of-plane component of the stray field $B_z(x,y)$.[27]

To help inform what kind of pattern of stray-field distribution one anticipates to observe, we show the simulated $B_z(x,y)$ generated by a square magnetic structure with a uniform out-of-plane (**Figure 2**a) and in-plane 2D magnetization (**Figure 2**b). The former generates a $B_z(x,y)$ that switches sign at the boundary, where the latter generates a dipole-like pattern that is antisymmetric along the magnetization direction. In **Figures 2**c-h, we show images of experimentally measured stray field distribution of several flakes. Thickness of flakes was measured with atomic force microscopy (AFM) and is 15.7(3) nm (16 layers) for **Figures 2**c, f, 28.6(3) (29 layers) for **Figures 2**d, g, and 25.6(2) nm (26 layers) for **Figures 2**e, h. Thickness is 56.5(6) nm (58 layers) for **Figure 2**i. In **Figures 2**c-e, we show $B_z(x,y)$ with an out-of-plane bias field $B_0$~30 mT. In each case, we observed $B_z(x,y)$ that has similar topology to the one shown in **Figure 2**a, indicating that magnetization is directed out-of-plane. We then lowered the bias field to $B_0$~3 mT. The corresponding $B_z(x,y)$, shown in **Figures 2**f-h, undergoes a dramatic change in topology and reveals a texture in stray field and hence underlying magnetic domains. The stray field texture contains multiple dipole-like features, consistent with the expectation of in-plane magnetized domains inferred from bulk magnetization measurement that shows zero remanence and easy-plane anisotropy. The change in stray field topology as the external field was lowered and the revelation of domains confirm ferromagnetic ordering of flakes; if the flakes were paramagnetic, the stray field would merely change in amplitude but not topology. In **Figure 2**i, we show the stray field of another flake at bias field ~3 mT, which also displays stray field texture, demonstrating that room-temperature ferromagnetism is ubiquitous among CFGT flakes.

### 2.3. Thickness-dependent $T_c$

We then measured $T_c$ of various flakes shown in **Figure 2**. We employed simultaneous magnetometry-thermometry as described in Ref. [27]. Heating was provided by a resistive heater, and ODMR imaging provides both $B_z$ and temperature $T$ of the sample. An out-of-plane bias field ~3 mT was used. Because of the presence of domains, we adopted a statistical technique to extract $T_c$[36]. For each flake, we select a region-of-interest (ROI) in the neighborhood of the flake, and compute the variance $\Delta B_{ROI}^2 = \langle (B_z - \langle B_z \rangle)^2 \rangle$, where $\langle ... \rangle$ denotes average over the ROI. Below $T_c$, the presence of stray field texture is captured in the variance, and hence one expects the variance to drop to zero at $T_c$ as temperature is raised, and remains zero beyond $T_c$. Since fluctuations in measurement can also contribute to variance, we also select a background region in the vicinity of the ROI, and compute the background variance $\Delta B_{BG}^2$. We then compute the normalized variance $\Delta B^2 = \Delta B_{ROI}^2 - \Delta B_{BG}^2$ which provides a proxy for magnetic ordering. In **Figures 3**a-c, $\Delta B^2$ vs $T$ is shown for the three flakes corresponding to **Figures 2**c-h. In general, we observe $\Delta B^2$ decreases gradually as $T$ increases, and then become flat and vanishingly small above a cutoff temperature, consistent with the expectation of a phase transition leading to the disappearance of ferromagnetic order. A fit to the power-law behavior $y_0+a(1-T/T_c)^\beta$ enables the extraction of $T_c$. We obtain $T_c$ = 309.8(7) K, 316(2) K, and 321(2) K for the 16, 26, and 29 layer flakes respectively.

Our result adds to the portfolio of exfoliated ultrathin 2D magnets at room-temperature, the other two being $CrTe_2$[22,23] and $VSe_2$ [24], which as discussed previously have potential robustness issues. In the context of 2D magnets with high $T_c$ (>250 K) created from exfoliaton of robust crystals, i.e. those that are thermally stable and do not undergo a structural transition when thinned down, a thickness-independent $T_c$ ~270K was found for $Fe_4GeTe_2$ down to 7 nm[25], while $T_c$ ~270 K for 12 nm $Fe_5GeTe_2$[26] and ~300 K for 20 nm $Fe_5GeTe_2$[27]. Hence, among materials exfoliated from robust crystals, our result of above-room-temperature ferromagnetism in ~16 nm thick flake corresponds to the thinnest room-temperature 2D magnet flake as well as the highest $T_c$ for ultrathin flake (<20 nm). This thickness enters the regime where practical spintronic devices can be created. For example, spin-torque oscillator has been fabricated from 17 nm thick yttrium iron garnet[37], while highly efficient spin-orbit-torque (SOT) switching has been demonstrated for ~15-20 nm thick $Fe_3GeTe_2$[13]. Hence, our results pave the way for the development of practical ultrathin devices based on 2D magnets operational at room-temperature.

For the thicker flake shown in **Figure 2**i (thickness ~56 nm corresponding to 58 layers), we observe nonhomogeneous temperature-dependent trends in different regions. For example, the region shown in **Figure 3**d is well described by a single power-law behavior with $T_c$ =325.9(3) K, while **Figure 3**e displays a bimodal behavior whose fit requires a superposition of two power-law terms of the form $y_0+a(1-T/T_c)^\beta$, giving rise to two transition temperatures ~316 and ~332 K. Another intriguing behavior is observed in the vicinity of the crack indicated in the ROI of **Figure 3**f. In **Figure 2**i, a large stray field is observed in the same region, which in fact persists to the highest temperature explored in this work. In **Figure 3**f, we plot the average stray field of this ROI in the vicinity of the crack. For comparison, we also plot the average stray field of a background indicated in the inset, which is nearly zero across all temperatures as expected. However, significant stray field associated with this ROI is observed up to ~350 K, indicating magnetization persists up to this temperature. As observed from AFM image (**Figure S3, Supporting Information**), this flake has several folds and faults. It is reasonable to expect that large and complex strain environment exists within this flake, leading to widely varied transition behavior. Our result suggests the possibility to tune and even enhance $T_c$ in CFGT via strain engineering[38].

## 2.4. Monte Carlo simulations of CFGT monolayer using *ab initio* effective spin Hamiltonian

To investigate $T_c$ in CFGT monolayer, we employed Monte Carlo simulation using *ab initio* effective spin Hamiltonian. The classical effective low-energy spin Hamiltonian[39,40] describing the magnetic states at relevant experimental energy scales

$$H = -\frac{1}{2}\sum_{i,j} \mathbf{S}_i \mathbf{J}_{ij} \mathbf{S}_j - K\sum_i (\mathbf{S}_i \cdot \hat{z})^2 - \mu_S \sum_i \mathbf{S}_i \cdot \mathbf{B}, \quad (1)$$

is constructed *ab initio* [39] using using noncollinear density functional theory (DFT) calculations. Here, $\mathbf{S}_i$ are unit vectors; $\mathbf{J}_{ij}$ is 3 × 3 tensor[40,41] describing isotropic nearest-neighbor and beyond (as denoted by $i, j$) exchange interaction, $J_{ij} = \frac{1}{3}\text{Tr}\mathbf{J}_{ij}$; the magnitude of the easy-plane anisotropy is specified by $K$; and $\mathbf{B}$ is an external magnetic field. Table 1 provides *ab initio* extracted $J_{ij}$ parameters between atoms of CFGT units cell depicted in **Figure 4**a. We also compute the out-of-plane anisotropy, $A = E_\perp - E_\parallel$, where $E_\perp$ and $E_\parallel$ are ground state energies for atomic magnetic moments aligned out-of-plane or in-plane, respectively, for both FGT and CFGT monolayers. Such anisotropy is $A = -1.5$ μeV/Fe-atom for FGT and $A = 2$ μeV/Fe-atom for CFGT monolayer. Thus, Co-substitution switches easy-axis anisotropy of FGT to easy-plane anisotropy of CFGT, as encoded by the second term in **Equation 1**. These values are in good agreement with the previous DFT calculations[29]. As both 20% and 28% Co-substituted FGT host easy-plane anisotropy, we extract $J_{ij}$ and $K$ parameters in **Equation 1** for 20% Co-substituted FGT to avoid computational expense due to large number of atoms in the supercell. Additional details can be found in **Supporting Information Section S1**.

At $T=0$, the calculated ground state of CFGT monolayer is an easy-plane ferromagnet, which becomes magnetized out-of-plane with a perpendicular external field at 40 mT. Normalized magnetization $m/m_s$ vs $T$ is shown in **Figure 4**b. A fit to $y_0 + a(1-T/T_c)^\beta$ yields $T_c = 271.6(9)$ K and $\beta = 0.340(6)$ (see **Section S2** in **Supporting Information** for details of the fit). The extracted $\beta$ corresponds very well to various 3D universality classes, including 3D XY ($\beta$ =0.345), 3D Heisenberg ($\beta$=0.365), and 3D Ising ($\beta$=0.325); in contrast, it does not correspond well to 2D XY ($\beta=\frac{3\pi^2}{128}\approx0.231$)[7,42] or 2D Ising ($\beta$=0.125) universality classes[7]. While nonzero magnetization of pristine monolayer FGT is made possible by its easy-axis anisotropy[43], as is the case of a number of 2D magnetic monolayers[3-6], CFGT has easy-plane anisotropy. Hence for CFGT, one may expect a scenario in which rotational freedom of magnetization in the plane prohibits long-range magnetic order according to the Mermin-Wagner theorem in the strictly 2D scenario and thermodynamic limit[44]. Nevertheless, it is recognized that finite size effects and large enough in-plane anisotropy do allow for nonzero magnetization in the Berezinskii-Kosterlitz-Thouless (BKT) scenario[43], where magnetization vanishes with the critical exponent $\beta=\frac{3\pi^2}{128}\approx0.231$[42] while concurrently decaying with the system size (albeit very slowly)[43]. While this 2D XY exponent has been very recently experimentally confirmed for monolayer $CrCl_3$[7], it appears to be not realized in monolayer CFGT, suggesting that it is not a true 2D XY system and hence not subject to the Mermin-Wagner theorem. One way to evade the Mermin-Wagner theorem is long-ranged exchange interaction, scaling with distance $r_{ij}$ between two spins as $J_{ij}\sim1/r_{ij}^p$ with $2<p<4$ required in 2D[45]; for our case using nearest-neighbor (NN) and next-NN exchange interaction in **Table 1**, this is insufficient [46] on its own. Instead, the supercell in **Figure 4**a shows two atomic planes where strong exchange interaction (**Table 1**) between them is likely a reason for evading both the Mermin-Wagner theorem and the BKT scenario[43].

We can compare the $T_c$ of CFGT to the result of pure FGT (lattice structure shown in **Figure 4**a, interaction parameters shown in **Table 2**) where we extracted $T_c$ =281.0(5) K (**Figure 4**b). The existence of a higher theoretical $T_c$ in the pure FGT system than in the Co-substituted system is consistent with the energetics of previous DFT calculations for these systems[29]. We speculate the following two reasons for higher experimental $T_c$ values in the Co-substituted samples: first, a larger total occupancy of the Fe1 sublattice and second, a reduction in the dynamics of the Fe1 sublattice magnetism due to chemical and/or short-range structural effects. Stacking faults and related strains may also need to be considered because they are known to impact the magnetism of this family[26], though likely are less relevant when considering a monolayer. These effects, unaccounted for by the simulation, may lead to a deviation of experimental monolayer $T_c$ from the value we extracted. Nevertheless, similar Monte Carlo calculations for $CrI_3$, $CrBr_3$, and $CrGeTe_3$ have yielded monolayer $T_c$ of 69, 39, and 65 K, which have reasonably good agreement (to within 25 K) with experimental values
of 45, 34, and 45 K respectively[47], and show the same qualitative order in terms of which
material has higher/lower $T_c$. Hence, while the simulations reported here are idealized
calculations in comparison to the complex experimental systems, the qualitative trends
reported by the simulations demonstrate that high $T_c$ may be expected for monolayer
materials. Future investigations are needed to understand how the transition metal content
controls the properties in this material.

While the calculated monolayer $T_c$ for CFGT is below room temperature, it is in fact not too
far from room temperature and even higher than freezer temperature (255 K). Hence, a
monolayer with this $T_c$ can readily operate outside of a cryostat when cooled with a
thermoelectric cooler (TEC). We also note that the calculation here only takes into account of
short-range interaction, and hence should be considered as a lower bound of $T_c$. Dipolar
interaction can further stabilize magnetic order, and hence one can expect an actual $T_c$ that is
higher than the one calculated here. In **Figure 4**c, we plot $T_c$ from experimental data as well
as from the monolayer calculation. An extrapolation estimates $T_c$ ~280 K from experimental
data, which is above but not too far from the calculated value. Hence, we surmise the actual $T_c$
of CFGT monolayer may very well lie in the range ~270-280 K. $T_c$ could be further enhanced
via micropatterning[48], strain engineering[38], or via proximity effect[49,50], all of which
have demonstrated >30 K enhancement in other 2D magnets. Hence, our result shows that
room-temperature or nearly-room-temperature device involving monolayer is now within the
realm of possibility, where at most a moderate cooling (e.g. via TEC) that is compatible with
a practical working environment is sufficient to ensure operation of the device.

## 3. Conclusion

In conclusion, we have studied exfoliated flakes of CFGT via QMI. We have observed above-room-temperature ferromagnetism in flakes as thin as 16 nm, with $T_c$ ranging from ~310 K in
the thinnest flake studied to ~330 K in the bulk. Our results correspond to one of the thinnest
room-temperature 2D magnets created from exfoliation, as well as the highest $T_c$ in ultrathin
(<20 nm) 2D magnetic flakes created via exfoliation of robust crystals that are thermally
stable and do not undergo structural transition when thinned down. These results show that
CFGT provides a suitable material platform for fabricating high quality devices. Via atomistic
*ab initio* calculation, we have obtained a $T_c$~270 K in CFGT monolayer. This work
establishes CFGT as a platform for high-quality 2D magnetic devices with high Curie temperature, and therefore opens up a number of important directions in the field of 2D magnets. First, our results demonstrate the possibility of monolayer device capable of operation outside of a cryostat requiring only moderate cooling (e.g. via TEC) as well as engineering room-temperature $T_c$ in monolayer via strain, micro-patterning, or proximity effect. Second, even without going into monolayer, our results show that room-temperature 2D magnet now enters the thickness regime in which practical devices can be made and that can take advantage of the unique properties of vdW materials. For example, excellent interface resulting from atomically smooth flat surface, which is a characteristic of vdW materials, can lead to highly-efficient SOT switching, even for flakes ~15-20 nm thick.[13] Hence, our results pave the way for creating energy-efficient spintronic devices at room-temperature with ultrathin CFGT flakes, such as SOT-magnetic random-access memory and spin-torque oscillator. Lastly, our work also opens up the direction towards engineering novel magnetic phases in ultrathin 2D magnet at or near room-temperature, e.g. non-collinear texture or skyrmion via proximity effect or twist engineering[8-10,51].

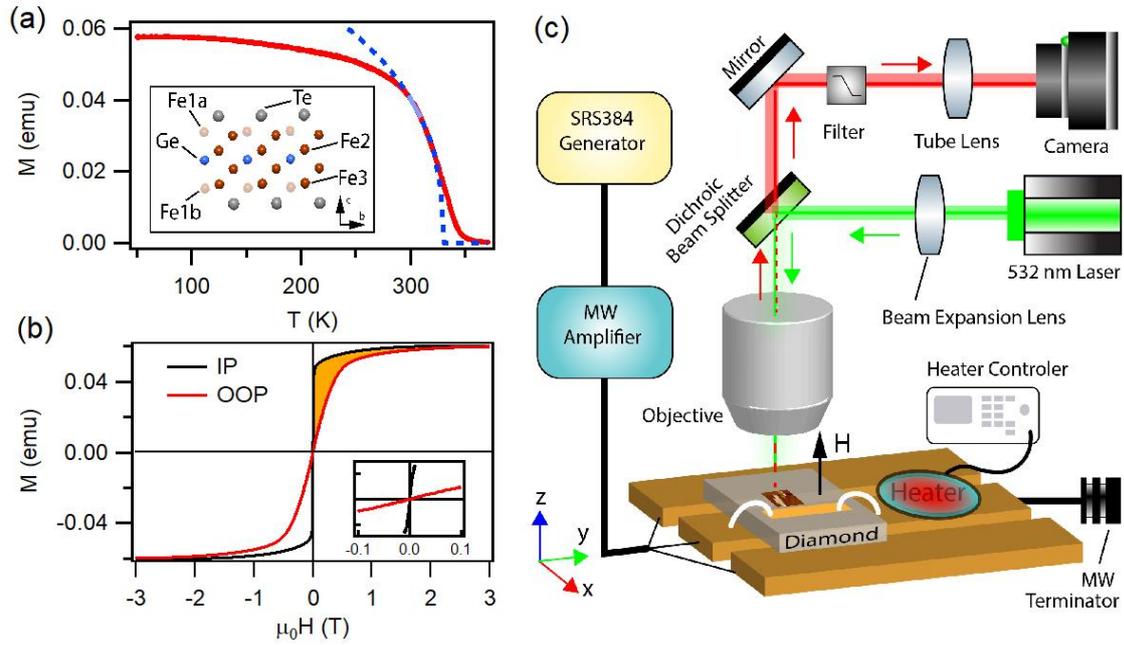

**Figure 1.** (a) Temperature-dependent magnetization measurement *M* vs *T* for bulk CFGT crystal. A fit near the critical region extracts $T_c$=328 K. The inset shows the lattice structure of CFGT; a single layer is shown. Fe1a and Fe1b correspond to a split site that allows for local atomic order and disorder; Co-substitution of these sites is energetically favorable compared to Fe2 or Fe3 sites. For simplicity, the associated split site of Ge is not illustrated. (b) Isothermal magnetization measurement *M* vs *H* measurement for in-plane (IP) and out-of-plane (OOP) *H*, showing saturation behavior and easy-plane anisotropy. For OOP data, the horizontal axis takes into account the demagnetization correction (see **Section S5** in **Supporting Information**). Inset: zoom-in bulk in-plane measurement shows zero remanence, and hence the presence of domains. (c) Schematics of QMI.

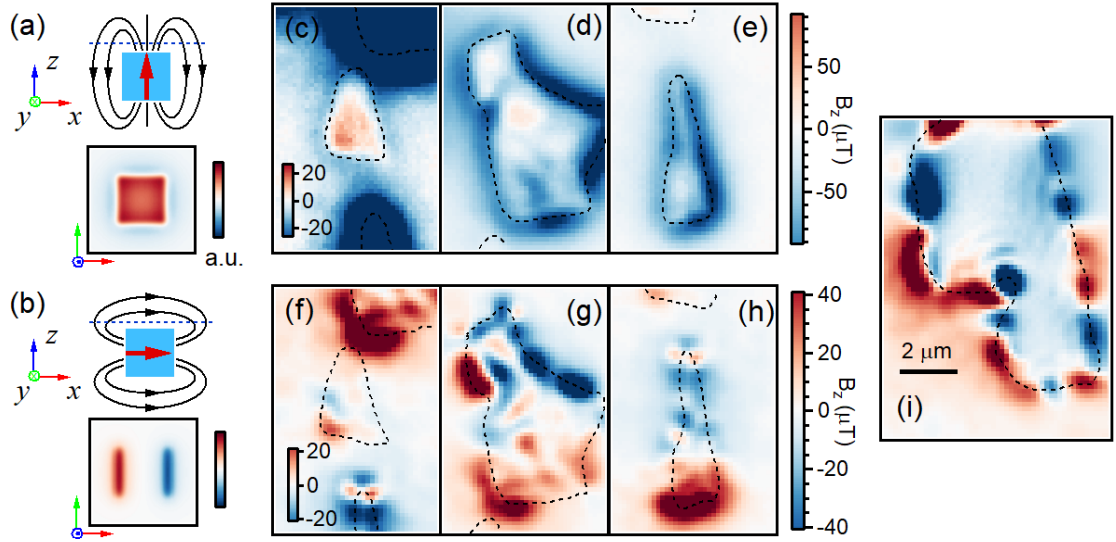

**Figure 2.** Quantum magnetic imaging of CFGT flakes. a) and b): Simulated distribution of vertical stray field $B_z$ for uniform a) out-of-plane (pointing along z) and b) in-plane (pointing along x) magnetization. Here, one assumes a 2D 1 μm × 1 μm square of magnetization, and stray field is simulated at a stand-off distance $d$=100 nm. Images are convolved with a point spread function corresponding to 600 nm optical resolution of the experimental setup. c)-e) Experimental stray field of CFGT flakes at room temperature with an applied vertical bias field (along $z$-direction) ~30 mT. These images have similar topology to a), demonstrating that magnetization is aligned along the vertical (z-) direction. f)-h): Measurement of the same flakes corresponding to c-e respectively, but with a smaller vertical bias field ~3 mT. Here, we see a stray field texture develops that reveals underlying domains. Magnetizations that get oriented along the bias field at high field and forms domains at low field demonstrate that these flakes have ferromagnetic order. Thickness of flakes are 15.7(3) nm (16 layers) for c, f, 25.6(2) nm (26 layers) for d, g, and 28.6(3) nm (29 layers) for e, h. i) Measurement of a 56.5(6) nm thick (58 layers) flake under the same condition as f-h. All experimental images are shown to the same scale; a 2 μm scale bar is shown in (i).

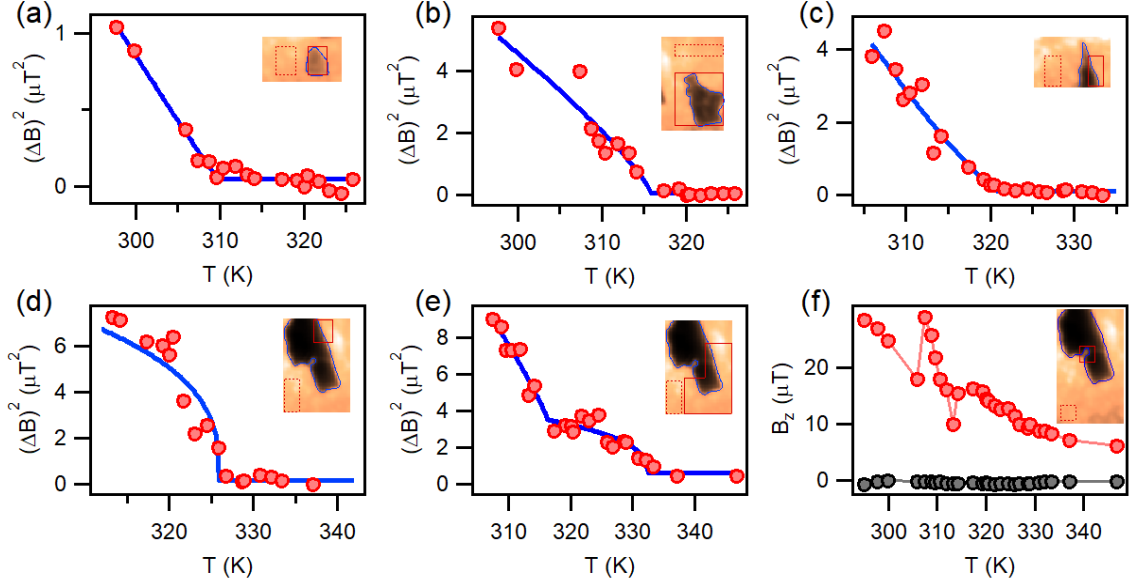

**Figure 3.** Temperature-dependent measurement. Panels (a-e) show normalized variance $\Delta B^2 = \Delta B_{ROI}^2 - \Delta B_{BG}^2$ as a function of temperature $T$, where $\Delta B_A^2 \equiv \langle (B_z - \langle B_z \rangle)^2 \rangle$ with $A$ corresponding to the area over which statistical averaging $\langle \ldots \rangle$ is performed. In the inset of each panel, an optical image of the flake is shown, together with the region-of-interest (ROI) in the neighborhood of the flake shown as solid red rectangle, and background (BG) shown as dashed red rectangle. Flake thickness is (a) 15.7 nm (6 layers), (b) 25.6nm (26 layers), (c) 28.6 nm (29 layers), and (d-e) 56.5 nm (58 layers). Experimental data of $\Delta B^2$ vs $T$ is shown in red circles. Data is fit to power-law behavior $y_0 + A(1 - T/T_c)^\beta$, except for (e) which is fit to a superposition of two such terms. Fit is shown in blue curve. Extracted $T_c$ is (a) 309.8(7) K, (b) 316(2) K, (c) 321(2) K, and (d) 325.9(3) K. For (e), two temperatures ~316 and ~322 K are extracted. (f) Average stray field of the ROI (red) and average stray field of the BG (black circles) vs $T$.

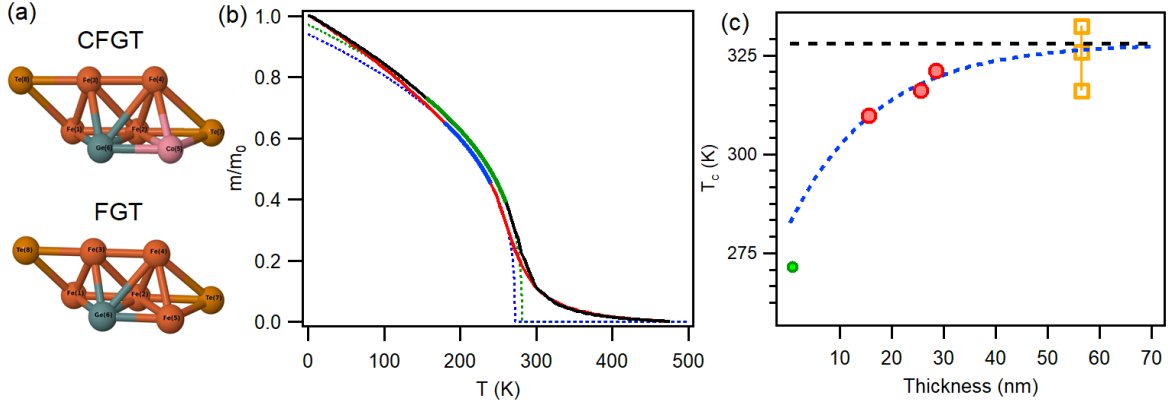

**Figure 4.** Monte Carlo simulation of CFGT and FGT monolayer. (a) Lattice considered in the calculation for CFGT (top) and FGT (bottom). Exchange couplings between spins are listed in Table 1 (CFGT) and Table 2 (FGT). (b) Average magnetization $m$ vs temperature $T$, normalized by magnetization $m_0$ at $T=0$. A fit to power-law $y_0+A(1-T/T_c)^\beta$ near the phase transition extracts $T_c$ =271.6(9) K and $\beta$=0.340(6) for CFGT, and $T_c$ =281.0(5) K and $\beta$=0.352(3) for FGT. Red (black) is simulation, blue (green) solid curve is the fitted model across the fitted temperature range, and blue (green) dashed curve is the fitted model across the entire temperature range for CFGT (FGT). (c) Experimental $T_c$ vs thickness for the 16-29 nm thick flakes (red) and 57 nm thick flake (orange), and CFGT monolayer value from calculation (green). Black dashed line is bulk value 328 K. Blue dashed curve: extrapolation via an exponential fit to red circles.

**Table 1.** Values of exchange coupling $J_{ij}$ between sites as labeled in **Figure 4**a for CFGT.

| System | $J_{ij}$ (meV) |
|---|---|
| Fe(1)-Fe(2) | 54.1539 |
| Fe(1)-Fe(3) | 33.5254 |
| Fe(1)-Fe(4) | -0.7548 |
| Fe(2)-Fe(3) | 2.1669 |
| Fe(2)-Fe(4) | 26.9815 |
| Fe(2)-Co(5) | 1.9647 |
| Fe(3)-Fe(4) | 28.4089 |
| Fe(4)-Co(5) | 7.6406 |

**Table 2.** Values of exchange coupling $J_{ij}$ between sites as labeled in **Figure 4**a for FGT.

| System | $J_{ij}$ (meV) |
|---|---|
| Fe(1)-Fe(2) | 42.2640 |
| Fe(1)-Fe(3) | 32.5760 |
| Fe(1)-Fe(4) | 0.0773 |
| Fe(2)-Fe(3) | 4.7758 |
| Fe(2)-Fe(4) | 21.1770 |
| Fe(2)-Co(5) | 9.1585 |
| Fe(3)-Fe(4) | 21.2840 |
| Fe(4)-Co(5) | 8.7697 |

**Acknowledgements**

HC and SA contributed equally to this work. This research was partially supported by NSF through the University of Delaware Materials Research Science and Engineering Center DMR-2011824 Seed Award program. Bulk crystal synthesis and characterization (AFM, MAM) were supported by the U. S. Department of Energy, Office of Science, Basic Energy Sciences, Materials Sciences and Engineering Division. KD, VMLDPG, MBJ, and BKN were supported by NSF through the University of Delaware Materials Research Science and Engineering Center, DMR-2011824. D.J. acknowledges support from a seed grant from the National Science Foundation (NSF) supported University of Pennsylvania Materials Research Science and Engineering Center (MRSEC) (DMR-1720530). H.Z. was partially supported by the Vagelos Institute of Energy Science and Technology graduate fellowship. ZJC and KL were supported by the NSF (DMR-2005108).

**Conflict of Interests**

The authors declare no conflict of interests.

**Data Availability Statement**

The data that support the findings of this study are available from the corresponding author upon reasonable request

**Supporting Information**

**Above-room-temperature ferromagnetism in ultrathin van der Waals magnet**

Hang Chen[1], Shahidul Asif [2], Kapildeb Dolui[3], Yang Wang[1], Jeyson Tamara Isaza[1,4], V. M. L. Durga Prasad Goli[1], Matthew Whalen[1], Xinhao Wang[1], Zhijie Chen[5], Huiqin Zhang[6], Kai Liu[5], Deep Jariwala[6], M. Benjamin Jungfleisch[1], Chitraleema Chakraborty[1,2], Andrew F. May[7], Michael A. McGuire[7], Branislav K. Nikolic[1], John Q. Xiao[1], Mark J.H. Ku[1,2,†]

[1]*Department of Physics and Astronomy, University of Delaware, Newark, Delaware 19716, United States*
[2]*Department of Materials Science and Engineering, University of Delaware, Newark, Delaware 19716, United States*
[3]*Lomare Technologies Ltd., 6 London Street, London EC3R 7LP, United Kingdom*
[4]*Departamento de Física, Universidad Nacional de Colombia, Bogotá D.C., Colombia*
[5]*Department of Physics, Georgetown University, Washington, DC 20057, United States*
[6]*Department of Electrical and Systems Engineering, University of Pennsylvania, Philadelphia, Pennsylvania 19104, United States*
[7]*Materials Science and Technology Division, Oak Ridge National Laboratory, Oak Ridge, Tennessee 37831, United States*

† Email: mku@udel.edu

### Section S1.
### Phase transition of monolayer CFGT and FGT from classical Monte Carlo simulations using *ab initio* effective spin Hamiltonian

The critical temperature $T_c$ and exponent $\beta$ with which magnetization of CFGT monolayer is vanishing $M \sim (T - T_c)^\beta$ at $T_c$ are obtained via classical Monte Carlo simulations performed for a monolayer of 4500 atoms with periodic boundary conditions. The classical effective low-energy spin Hamiltonian[1] describing the magnetic states at relevant experimental energy scales

$$H = -\frac{1}{2}\sum_{i,j}\mathbf{S}_i \mathbf{J}_{ij} \mathbf{S}_j - K\sum_i (\mathbf{S}_i \cdot \hat{z})^2 - \mu_S \sum_i \mathbf{S}_i \cdot \mathbf{B}, \qquad (1)$$

is constructed *ab initio*[1a] using noncollinear density functional theory (DFT) calculations. In the extended classical Heisenberg Hamiltonian in **Equation S1**, atomic magnetic moments of magnitude $\mu_S$ are treated as classical vectors[1b, 2] whose magnitude remains fixed during their rotation, which is a good approximation when atomic magnetic moments are large enough. It can be used as a starting point of classical Monte Carlo or atomistic spin dynamics[2] simulations. Here
$S_i = \mu_S/|\mu_S|$ are unit vectors; $\mathbf{J}_{ij}$ is $3\times 3$ tensor[1b, 2] describing isotropic interactions $J_{ij} = \frac{1}{3}\mathrm{Tr}\mathbf{J}_{ij}$; the magnitude of the easy-plane anisotropy is specified by $K$; and $\mathbf{B}$ is an external magnetic field. Table 1 (Table 2) of the main text provides *ab initio* extracted $J_{ij}$ parameters between atoms of CFGT (FGT) units cell depicted in **Figure. 4**a of the main text.

We adopt the lattice structure provided in the file CalcRelaxed_Fe5GeTe2_Fe1a_only_AllUp_Structural.cif from Ref.[3]. This is the structure in which the Fe1 atoms occupy only the Fe1a position. This is the same model considered in Ref.[4]. DFT and experiment[3] indicate that disorders such as split sites and stacking faults are present in the real material and play a role in determining magnetic properties. Nevertheless, the ordered model used here serves as a conceptual model for the material to enable computation to be performed. Despite the idealized nature of the structure, we expect that the results obtained with this structure should provide a decent estimate that enables us to see qualitatively where $T_c$ of monolayer CFGT lies with respect to room temperature. Further confidence in the calculation is provided by 1) the agreement of calculated results from CFGT and FGT monolayer and multilayer with expected behaviors (**Section S3**), and 2) the decent match between the calculation and extrapolated $T_c$ of monolayer CFGT.

The parameters $J_{ij}$ are calculated using Green's function (GF) formalism of Ref. [5], as implemented in QuantumATK package[6], and given in **Table 1** (**Table 2**) between nearest-neighbors and next-nearest-neighbor for CFGT (FGT). The Kohn-Sham Hamiltonian of DFT, as the input of GF formalism, is obtained from noncollinear DFT calculations using the Perdew-Burke-Ernzerhof (PBE) parametrization[7] of the generalized gradient approximation (GGA) to the exchange-correlation functional, as implemented in QuantumATK package[6]; norm-conserving fully relativistic pseudopotentials of the type PseudoDojo-SO[6, 8] for describing electron-core interactions; and the Pseudojojo (medium) numerical linear combination of atomic orbitals (LCAO) basis set[8]. The energy mesh cutoff for the real-space grid is chosen as 151 Hartree, and the $k$-point grid 21×21×7 is used for the self-consistent calculations. As both 20% and 28% Co-substituted FGT host easy-plane anisotropy, we extract $J_{ij}$ and $K$ parameters in **Equation S1** for 20% Co-substituted FGT to avoid computational expense due to large number of atoms in the supercell.

### Section S2.
### *Extracting $T_c$ and critical exponent of data from Monte Carlo simulation of CFGT monolayer*

The power-law model involving normalized magnetization

$$m_n = (1-T/T_c)^\beta, \tag{S2}$$

where $m_n = m/m_0$ is magnetization $m$ normalized by zero-temperature magnetization $m_0$, allows for the extraction of the critical temperature $T_c$ and critical exponent $\beta$, but is only valid in the neighborhood $T_{start} < T < T_c$ of the critical point. Hence, we need to determine a suitable window for the fit. Also, due to critical slowing-down and finite computation resource, calculated $m$ will have a greater deviation from the true value right around the critical point. Hence, we need to find the optimal window $T_{start} < T < T_{end}$ (with $T_{end} < T_c$) for the fitting.

To determine $T_{end}$, we first note if the power-law model is valid, then

$$m_n/(dm_n/dT) = (T - T_c)/\beta, \tag{S3}$$

namely $m_n/(dm_n/dT)$ should be linear. In **Figure S1**a, we show $m_n/(dm_n/dT)$ vs $T$ and see that it is indeed linear around ~200-250 K. At around 240-250 K, it starts to bend around and then become somewhat flat, signaling a transition from ordered to paramagnetic phase. The transition is not completely sharp, which is likely due to critical slowing-down in the calculation. Nevertheless, if we look at the slope of $m_n/(dm_n/dT)$ vs $T$, we should see the slope displays a jump from a finite value to zero. In **Figure S1**b, we show $d[m_n/(dm_n/dT)]/dT$ vs $T$, and indeed we see a jump around ~240-250 K. $d[m_n/(dm_n/dT)]/dT$ vs $T$ is more or less flat until 240 K, and then drop to zero. Hence, $T_{end}$=240 K is used as the end-point of the fit.

Having determined the upper range of the fit, we then perform fit to the power-law model **Equation S2** over the range $T_{start} \leq T \leq T_{end}$ with different $T_{start}$. To determine the best $T_{start}$, we look at the uncertainty $\delta T_c$ of $T_c$ which provides a measure of the goodness of the fit. **Figure S1**c shows that $\delta T_c$ is minimized for $T_{start}$=180 K. Hence, we fit the $m_n$ vs $T$ data to **Equation S2** over the range $T_{start}$ =180 K$\leq T \leq T_{end}$=240 K, which yields $T_c$ =271.6(9) K and critical exponent $\beta$=0.340(6); **Figure 4**b of the main text compares the fit to the data. For a consistency check, the line given by **Equation S3** with extracted $T_c$ and $\beta$ is shown in **Figure S1**a, which displays excellent agreement with $m_n/(dm_n/dT)$ vs $T$ data over the fitted temperature range, consistent with the expectation of **Equation S3**. Likewise, in **Figure S1**b, we see the expected behavior that the extracted $T_c$ coincides with the vanishing of $d[m_n/(dm_n/dT)]/dT$.

To investigate the sensitivity of $T_c$ and $\beta$ to the fitting range, in **Figure S1**d we show extracted $T_c$ and $\beta$ over a range of $T_{start}$ (with fixed $T_{end}$=240 K) in which $\delta T_c$ <1 K (hence the fit can be considered to match well to the data). Both $T_c$ and $\beta$ are somewhat sensitive to the fitting range. Nevertheless, within the range of $T_{start}$ considered, $T_c$ varies by no more than 6 K (or 2% in terms of relative deviation). There is a larger variation for $\beta$, which ranges ~0.31 to 0.37 over the values of $T_{start}$ considered. The sensitivity of $T_c$ and $\beta$ to the fitting range provides the rationale for the unbiased fitting procedure described above. Even with this sensitivity, we see that $T_c$ does not vary much and hence confidence can be placed in the reported value of $T_c$. The relative variation of $\beta$ is larger, but nevertheless the range of values of $\beta$ agrees much better to 3D universality classes ($\beta$=0.325, 0.345, and 0.365 for Ising, XY, and Heisenberg respectively) instead of 2D ones ($\beta$=0.125 and 231 for Ising and Berezinskii-Kosterlitz-Thouless in XY respectively). This supports the analysis in the main text that monolayer CFGT behaves more like a 3D system than a strictly 2D system, which provides a reason why Mermin-Wagner theorem does not apply and the system exhibits long-range magnetic order.

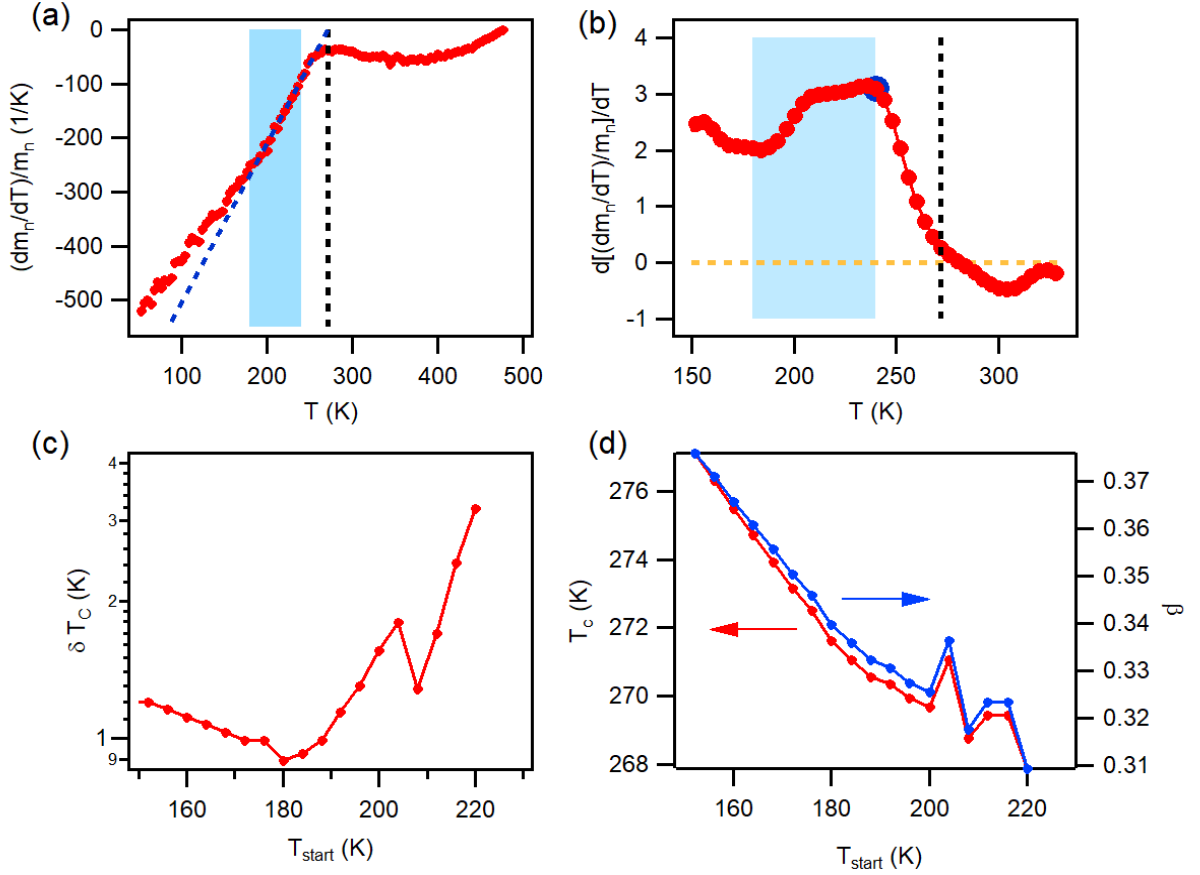

**Figure S1**. Analysis of calculated $m_n$ vs $T$ data for monolayer. (a) $m_n/(dm_n/dT)$ vs $T$ (red). (b) $d[m_n/(dm_n/dT)]/dT$ vs $T$ (red). As the raw derivative is noisy, a smoothening is applied. In (a) and (b), the blue band is the optimal region for fitting obtained from panel (c), and the black vertical dashed line is $T_c$ from fitting $m_n$ vs $T$ to the power-law model **Equation S1** over the optimal region. In (a), the blue dashed line is the line $(T-T_c)/\beta$. In (b), the orange dashed line is zero. (c) Fitting error on $T_c$ vs $T_{start}$. (d) $T_c$ (red, left vertical axis) and $\beta$ (red, right vertical axis) vs $T_{start}$.

### Section S3.
*Monte Carlo simulation results for CFGT and FGT*

Besides CFGT and FGT monolayer (ML), we have also performed Monte Carlo simulation for CFGT and FGT trilayer (TL) to check if $T_c$ follows an expected evolution with respect to layer number. The extraction of $T_c$ and $\beta$ follows the same protocol as described in **Section S2**. The results are summarized in **Table S1**. We observe the following trends which are all in agreement with expectations: 1) for a given layer number, $T_c$ is higher for FGT compared to CFGT (see the discussion in the main text on why this is the expected trend for theoretical calculations), and that 2) for a given material, $T_c$ is higher for multilayers compared to monolayer.

|  | $T_c$ (K) | $\beta$ |
|---|---|---|
| CFGT ML | 271.6(9) | 0.340(6) |
| FGT ML | 280.9(5) | 0.352(3) |
| CFGT TL | 307.8(5) | 0.476(2) |
| FGT TL | 314.9(7) | 0.424(4) |

Table S1: extracted $T_c$ and $\beta$ from Monte Carlo simulations.

### Section S4.
### Extracting $T_c$ of bulk crystal from magnetometry data.

In this section, we discuss the details of extracting $T_c$ from $M$ vs $T$ measurement of the bulk crystal (**Figure 1**a) obtained using vibration sample magnetometer (VSM). To do so, we fit $M$ vs $T$ to the power-law model **Equation S2**. Similar to **Section S2**, this model is valid over a range of temperature $T_{start} < T < T_{end}$ near the phase transition which we need to determine. However, since the VSM data is more noisy to differentiate compared to Monte Carlo simulation data, we use another approach to determine $T_{start}$ and $T_{end}$. Here, we simply fit the data to **Equation S2** over a range $T_{start} < T < T_{end}$, and we extract the parameters and their errorbars as we vary $T_{start}$ and $T_{end}$. To make the fit robust, we fix $\beta = 0.365$ corresponding to 3D Heisenberg universality class. This is appropriate given that the bulk crystal is known to have in-plane anisotropy (**Figure 1**b). **Figure S2** shows the errorbar on $T_c$ vs $T_{start}$ and $T_{end}$. An optimal range is found for $T_{start}=298.04$ K and $T_{end}=310.22$ K. Fitting **Equation S2** (with $\beta = 0.365$) yields $T_c = 328.45(3)$ K (**Figure 1**b).

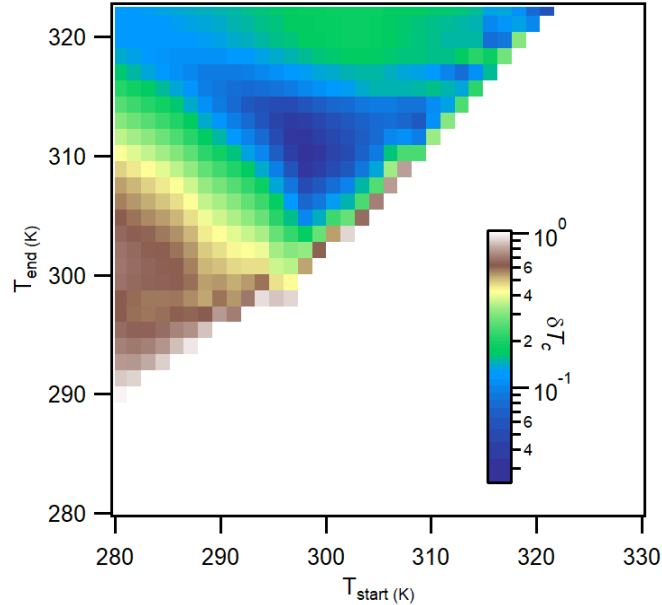

**Figure S2**. Determination of the optimal fitting range for VSM data $M$ vs $T$ of the bulk crystal. Shown is errorbar $\delta T_c$ on $T_c$ from the fit of the $M$ vs $T$ to the power-law model **Equation S2** over a temperature range $T_{start} < T < T_{end}$. $\delta T_c$ is determined as function of $T_{start}$ and $T_{end}$. Here, we fix $\beta = 0.365$. $\delta T_c$ is minimized at $T_{start}=298.04$ K and $T_{end}=310.22$ K.

### Section 5.

**Extracting anisotropy constant $H_a$ of bulk crystal from magnetomery data.**

The *M-H* curves in **Figure 1**b indicate in-plane anisotropy in bulk crystal. It is noted that in **Figure 1**b, we plot isothermal magnetization as a function of internal field $H_{in}$ after estimating demagnetization effect. For external field $H \parallel c$ axis, $H_{in} = H - NM$, where $N$ is the demagnetization factor and $M$ is the magnetization. Here, $N = 0.9$ is used to correctly estimate the $H_{in}$[3]. The effective anisotropic field $H_a$ can be calculated by $H_a = -2 K_{eff}/\mu_0 M_s$, where $K_{eff}$ is the effective magnetic anisotropic energy, $\mu_0$ is vacuum permeability and $M_s$ is the saturated magnetization[9]. Assuming $V$ is the volume of the sample, $H_a = -2(K_{eff} V)/\mu_0(M_s V) = -2A/\mu_0 m_s$, where $m_s$ is the total magnetic moment of the sample and $K_{eff}$ the total anisotropic energy, and $A$ equals to the area enclosed between the in-plane (IP) and out-of-plane (OOP) *M-H* curves. In addition, in the cgs system the $\mu_0$ is dimensionless and $\mu_0=1$ in our calculation. In **Figure 1**b, the integral over the enclosed area gives $A=170$ emu·Oe while the $m_s = 0.0668$ emu is obtained at $\mu_0 H=2.8$ T. As a result, $\mu_0 H_a = -0.51$ T.

**Section 6.**
**Atomic force microscopy image of thick CFGT flake**

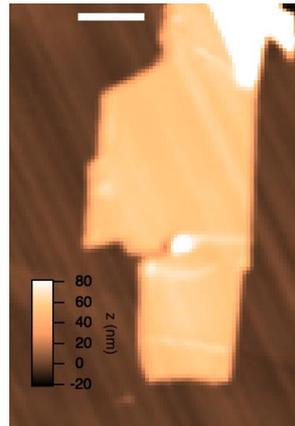

**Figure S3.** Atomic force microscopy (AFM) image of the thick CFGT flake in **Figure 2**i and **Figure 3**d-f. Scale bar is for 2 μm. Diagonal stripes in the background correspond to polished diamond surface. Multiple faults can be observed in this particular flake. Furthermore, the flake is bent and folded partially towards the top-right corner of the image.